%% file: manuscript.tex
\documentclass{sig-alternate}
\usepackage{algorithm}
\usepackage{algpseudocode}
\usepackage{subfigure}
\usepackage{psfrag}
\usepackage{cite}
\usepackage{balance}

\newtheorem{example}{Example}

\begin{document}
%
% --- Author Metadata here ---
%\conferenceinfo{DAC}{'09 San Francisco, California USA}
%\CopyrightYear{2007} % Allows default copyright year (200X) to be over-ridden - IF NEED BE.
%\crdata{0-12345-67-8/90/01}  % Allows default copyright data (0-89791-88-6/97/05) to be over-ridden - IF NEED BE.
% --- End of Author Metadata ---

%\title{Pessimism Reduction in Static Noise Analysis by Exploiting Logical and Temporal Correlation}
\title{FRAME: Fast and Realistic Attacker Modeling and Evaluation for Temporal Logical Correlation in Static Noise}

%\subtitle{[Extended Abstract]
%\titlenote{A full version of this paper is available as
%\textit{Author's Guide to Preparing ACM SIG Proceedings Using
%\LaTeX$2_\epsilon$\ and BibTeX} at
%\texttt{www.acm.org/eaddress.htm}}

\numberofauthors{5} %  in this sample file, there are a *total*
\author{
% 1st. author
\alignauthor
Sungroh Yoon\\
%       \affaddr{School of EE}\\
       \affaddr{Stanford University}\\
       \affaddr{Stanford, CA 94305, USA}\\
       \email{\textsf{sryoon@gmail.com}}
% 2nd. author
\alignauthor
Nahmsuk Oh\\
       \affaddr{Synopsys Inc.}\\
%       \affaddr{Camarillo}\\
       \affaddr{CA 93012, USA}\\
       \email{\textsf{Nahmsuk.Oh@synopsys.com}}
% 3rd. author
\alignauthor Peivand Tehrani\\
       \affaddr{Synopsys Inc.}\\
%       \affaddr{Camarillo}\\
       \affaddr{CA 93012, USA}\\
       \email{\textsf{Peivand.Tehrani@synopsys.com}}
\and  % use '\and' if you need 'another row' of author names
% 4th. author
\alignauthor Eui-Young Chung\\
%       \affaddr{Department of EE}\\
       \affaddr{Yonsei University}\\
       \affaddr{Seoul 120-749, Korea}\\
       \email{\textsf{eychung@yonsei.ac.kr}}
\alignauthor
Giovanni De Micheli\\
%       \affaddr{Centre for Integrated Systems}\\
       \affaddr{EPFL}\\
       \affaddr{Lausanne, Switzerland}\\
       \email{\textsf{giovanni.demicheli@epfl.ch}}
}
% There's nothing stopping you putting the seventh, eighth, etc.
% author on the opening page (as the 'third row') but we ask,
% for aesthetic reasons that you place these 'additional authors'
% in the \additional authors block, viz.
%\additionalauthors{Additional authors: John Smith (The Th{\o}rv{\"a}ld Group,
%email: {\texttt{jsmith@affiliation.org}}) and Julius P.~Kumquat
%(The Kumquat Consortium, email: {\texttt{jpkumquat@consortium.net}}).}
% Just remember to make sure that the TOTAL number of authors
% is the number that will appear on the first page PLUS the
% number that will appear in the \additionalauthors section.

\maketitle
\begin{abstract}
We propose a method called \emph{Fast and Realistic Attacker Modeling and Evaluation} (FRAME) that can reduce pessimism in static noise analysis by exploiting temporal logical correlation of attackers and using novel techniques termed \emph{envelopes} and \emph{$\sigma$ functions}. Unlike conventional pruning-based approaches, FRAME efficiently considers all relevant attackers, thereby producing more realistic results. FRAME was tested with complex industrial design and successfully reduced the pessimism of conventional techniques by 30.4\% on average, with little computational overhead. %We believe that FRAME can be a very fast and effective tool for pre-screening noise violators.
%This method is based upon modeling attacker and victim waveforms using triangular waves, from which we derive a combined effect of the attackers on the victim, considering both logic and temporal correlations among them. Our focus is on speed with little sacrifice to accuracy. (More to follow..)
\end{abstract}

%good expression: ``less pessimistic but still conservative'', ``prevents optimism and reduces unnecessary pessimism''

%A category including the fourth, optional field follows...
%\category{D.2.8}{Software Engineering}{Metrics}[complexity measures, performance measures]
%\category{B.6}{Logic Design}
%\category{B.6.3}{Logic Design}{Design Aids}%[simulation]
%\categroy{C.5.4}{Computer Systems Organization}{Computer System Implementation}[VLSI systems]
%\category{J.6}{Computer-Aided Engineering}{Computer-aided design (CAD)}



\section{Introduction}
Addressing noise issues in digital circuit design has become critical, due to the combined effects of technology scaling, high clock frequencies, and increased use of noise-sensitive dynamic circuits to fulfill high-speed requirements. Without thorough verification, noise effects will degrade the functionality, performance and reliability of a circuit. Most industrial static noise analyzers assume that all attackers switch simultaneously in the same direction, which frequently produces unrealistic results. Thus, pessimism reduction in static noise analysis has gained much attention~\cite{chen99, becer05, glebov04, das08, palla06, tseng05, ran05, glebov02, yang03, lee04, glebov02acm, palla08,arun01,xiao01,chen01,xiao00,xiao00b,chai03}.

%%%%%%%%%%%%%%%%%%%%%%%%%%%%%%%%%%%%%%%%%%%%%%%%%%%%%%%%%%%%%%%%%%%%%%
% complete references
%~\cite{chen99, becer05, glebov04, das08, palla06, tseng05, ran05, glebov02, yang03, lee04, glebov02acm, palla08,arun01,xiao01,chen01,xiao00,xiao00b,chai03}.
%%%%%%%%%%%%%%%%%%%%%%%%%%%%%%%%%%%%%%%%%%%%%%%%%%%%%%%%%%%%%%%%%%%%%%

Most existing techniques exploit temporal and/or logical correlations among attackers. Early approaches~\cite{chen01,xiao00,xiao00b,becer05} focused on timing correlations, which can be estimated by propagating timing windows through \emph{static timing analysis} (STA). However, considering only timing correlations is insufficient. For instance, temporally uncorrelated attackers can still be under a logical constraint that prevents them from switching in the same direction. Moreover, overlooking the interplay between temporal and logical correlation could give non-conservative timing results~\cite{das08}. Consequently, combined approaches that consider both temporal and logical correlations step-by-step~\cite{chen99, glebov04, palla06, glebov02, yang03, lee04, glebov02acm, palla08,arun01,xiao01} or simultaneously~\cite{tseng05, ran05,das08,chai03} have been suggested.

%Addressing signal integrity issues in digital circuit design has become critical, due to the combined effects of technology scaling, high clock frequencies, and increased use of noise-sensitive dynamic circuits to fulfill high-speed requirements. Without thorough verification, signal integrity effects will degrade the functionality, performance and reliability of a circuit. In this paper, we are interested in capacitive crosstalk noise analysis, particulary in reducing pessi

%The proposed FRAME method differs from existing techniques in several ways. First,

Most previous approaches aim at either pruning out invalid attackers~\cite{yang03,ran05,lee04,chai03,arun01,tseng05,das08,palla08} or selecting appropriate attackers and constructing a set of valid attackers~\cite{glebov02, glebov02acm, glebov04}. The problem of handling temporal and logical correlations is in general NP-hard~\cite{chen99}, and these combinatorial approaches depend critically on various techniques for alleviating the hardness, such as SAT solvers~\cite{tseng05,palla08, ran05,chai03}, ATPG/path-sensitization~\cite{arun01,xiao01,lee04,das08}, ROBDDs~\cite{yang03,glebov02,glebov02acm}, and other heuristics~\cite{palla06}. Nevertheless, many techniques still suffer from limited scalability to large-scale circuitries. Moreover, not considering all attackers can sometimes increase pessimism, deviating from the common belief that pessimism can always be reduced by pruning attackers (e.g. see Section~\ref{s-exp}).

The proposed FRAME method is related to existing approaches in that temporal and logical correlations are exploited together. On the other hand, FRAME is unique in the following aspects: First, FRAME implicitly considers all attackers, rather than pruning out or selecting attackers. Second, FRAME exploits a simple, graphical representation of attacker waveforms for exploring the search space without explicit enumeration of all possibilities. Taken together, FRAME runs fast and produces realistic results since the effect of every relevant attacker is simultaneously considered.

In a verification flow, FRAME would be positioned as a first-level noise estimator that can quickly examine large circuits and identify noise issues for more accurate analysis. Toward that end, FRAME exploits novel techniques termed \emph{envelopes} and \emph{$\sigma$ functions}, which enable us to compute the worst-case combined waveform of multiple attackers and their alignment information in time polynomial to the number of attackers, under reasonable assumptions on attacker waveform shapes and secondary attacker effects.

%Based upon the worst-case magnitude FRAME returns, we can identify noisy nets that need further analysis. Using the alignment information, high-accuracy simulators can be invoked for those problematic nets.

We extensively tested FRAME with 201 circuits extracted from a millon-gate 90-nm industrial design and discovered that FRAME can reduce pessimism by 30.4\% on average, compared with conventional pruning-based techniques, with little computational overhead. Based on our promising results, we believe that FRAME can be a practical and powerful tool for facilitating the entire verification flow.

\section{Proposed Method}
\input{method}

\section{Experimental Results}\label{s-exp}
\input{exp}

%\balance
\section{Conclusion and Future Work}
We have proposed FRAME, a fast and realistic method for modeling and evaluating multiple attackers in static noise analysis. The key difference between FRAME and conventional methods is that FRAME does not prune attackers completely but estimates the effect of all attackers without much computational burden, leveraged by efficient tracking of the worst-case magnitude and alignment information using envelopes and $\sigma$ functions. According to our experiments, FRAME can reduce pessimism by 30.4\% on average compared with conventional pruning-based attacker filtering techniques. Based upon our results, we believe that FRAME would be a very effective tool for pessimism reduction in static noise analysis.

FRAME can be extended so that it can handle more complex attacker structures such as multiple attacker chains. For higher accuracy, it would be helpful to employ non-linear interpolation for envelope addition and to model waveform widths more delicately.

%ACKNOWLEDGMENTS are optional
%\section{Acknowledgments}
%The authors would like to thank the anonymous reviewers of this manuscript for their review efforts.
%The authors would like to thank Beomjoon Seo and Hanjoo Kim for their help in manuscript preparation.
%The authors would like to thank Gerald Murray of ACM for his help in codifying this \textit{Author's Guide} and the \textbf{.cls} and \textbf{.tex} files that it describes.

%\bibliographystyle{unsrt}
%\balance

%\scriptsize
%\input{reference}

\balance

\bibliographystyle{abbrv}
%\scriptsize
%\balancecolumns
\bibliography{xtalk}  % sigproc.bib is the name of the Bibliography in this case
% ACM needs 'a single self-contained file'!
%\balancecolumns

%\balancecolumns % GM June 2007
% That's all folks!
\end{document}

%% file: method.tex
\subsection{Attacker Modeling and Assumptions}\label{ss-model-corr}
\input{assumption}

%\section{Proposed Method}

\subsection{The Objective and Overview of FRAME}\label{ss-objective}
\input{objective}

\subsection{Tracing the Worst Case By Envelopes}\label{ss-env}
\input{envelope}

\subsection{Recording Shift Amount by $\sigma$ Function}\label{ss-sigma}
\input{sigma}

\subsection{Representing Envelopes and $\sigma$ Functions}\label{ss-add}
\input{manipulation}

\subsection{Calculating the Worst-case Waveform}\label{ss-worst}
\input{worst}

%\subsection{Implementation}\label{ss-implementation}
%\input{manipulation}

%\subsection{Extensions and Future Work}\label{ss-extension}

%% file: assumption.tex
%\subsection{Modeling Attacker Correlation}\label{ss-model-corr}

\begin{figure}%
\centering
\scriptsize
\psfrag{t}{$t$}
\psfrag{0}{$0$}
\subfigure[Breaking attacker chain] {
    \psfrag{b}{3 attackers}
    \psfrag{v}{victim}
    \psfrag{i}{\!\!\!\!\!\!\!\!\!\!\!\!\!attacker timing windows}
    \psfrag{d}{attacker bumps}
    \includegraphics[width=\linewidth]{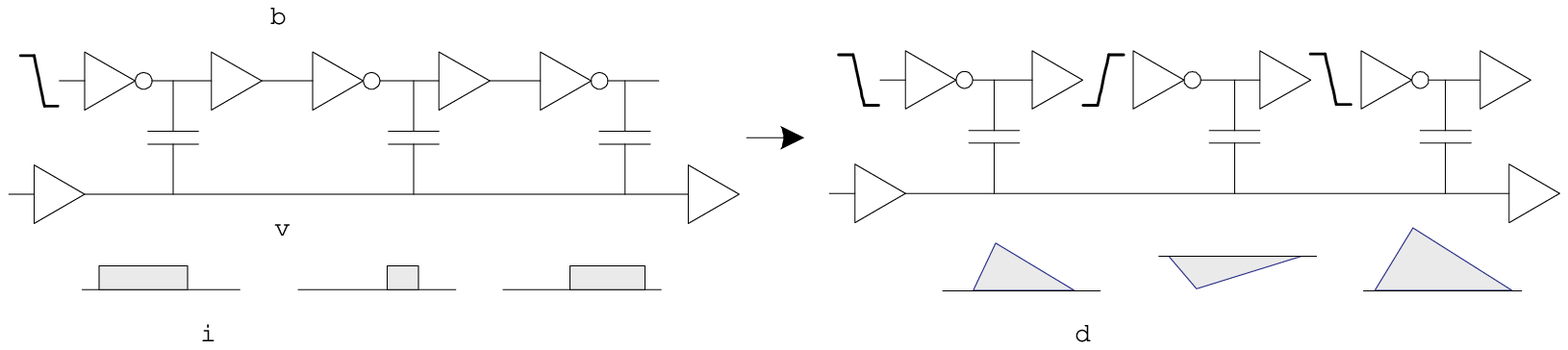}
    \label{f-chain}
}
\subfigure[Timing window] {
    \psfrag{a}{$a$}
    \psfrag{b}{$b$}
    \psfrag{x}{\!\!\!\!\!\!Earliest}
    \psfrag{y}{~~Latest}
    \psfrag{z}{\!\!\!\emph{timing window}}
    \includegraphics[width=0.5\linewidth]{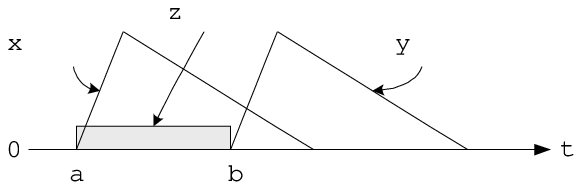}
    \label{f-timing-window}
}
%\subfigure[Triangularization: $\triangle(t;l,p,r,h)$] {
\subfigure[$\triangle(t;l,p,e,m)$] {
    \psfrag{l}{$l$}
    \psfrag{r}{$e$}
    \psfrag{h}{$\!\!|m|$}
    \psfrag{p}{$p$}
    \psfrag{x}{\!\!\!\!\!\!\!\!\!\!$\frac{m}{p-l}(t-l)$}
    \psfrag{y}{\!\!\!\!\!\!\!$\frac{-m}{e-p}(t-e)$}
    \includegraphics[width=0.4\linewidth]{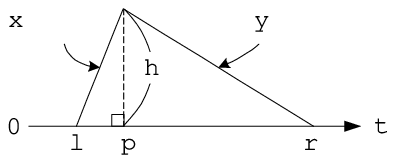}
    \label{f-triangle}
}
\caption{Attacker modeling.}
\end{figure}

Consider a chain of buffers and inverters whose nets (\emph{attackers}) are capacitively coupled with a common net (\emph{victim}), as shown in the left pane of Fig.~\ref{f-chain}. The input to the first gate in the chain is referred to as the \emph{chain head}. As the chain head (and its downstream gates thereafter) switches, the victim, which is assumed initially static, can be affected and dynamically change. We are interested in how individual attackers interact to create a composite waveform at the victim. Of particular interest is the worst-case combination that contains a peak of the largest magnitude.

%(For a discussion on a more general scenario, see Section~\ref{ss-extension}.)

Since the attackers on the chain are temporally and logically correlated, the worst-case search problem becomes computationally hard~\cite{chen99}. To alleviate that, we break the chain, apply separate input to each attacker block to induce an attacker bump, and estimate the original combined waveform using these bumps, as in Fig.~\ref{f-chain}. Temporal correlation among attackers is considered using timing windows, each of which is obtained through STA and provides the earliest and latest arrival time of an attacker bump, as shown in Fig.~\ref{f-timing-window}. Logical correlation is considered using the polarity of attacker bumps.
%
%In addition, FRAME assumes the following:
%\vspace{-0.5em}
%\begin{enumerate}
%\item[\textbf{A.1})] A single attacker waveform is piecewise linear and unimodal. It is thus modeled by a triangular wave.% as shown in Fig.~\ref{f-triangle}.
%\vspace{-0.5em}
%\item[\textbf{A.2})] The input slew of each attacker is fixed within its timing window.
%\vspace{-0.5em}
%\item[\textbf{A.3})] The effects of secondary attackers (i.e. an attacker of an attacker) are marginal and can be ignored.
%\end{enumerate}
%\vspace{-0.5em}

For computational efficiency, FRAME assumes the following: (\textbf{A.1}) A single attacker waveform is piecewise linear and unimodal. It is thus modeled by a triangular wave; (\textbf{A.2}) The input slew of each attacker is fixed within its timing window; (\textbf{A.3}) The effects of secondary attackers (i.e. an attacker of an attacker) are marginal and can be ignored.

Based on (\textbf{A.1-2}), we model each attacker bump by a triangle with four stationary parameters, as shown in Fig.~\ref{f-triangle}:
\vspace{-1em}
\begin{equation}\label{e-tri}
\triangle(t; l, p, e, m) =
    \begin{cases}
    \frac{m}{p-l}(t-l), &\text{if $l \le t < p$;}\\
    \frac{-m}{e-p}(t-e), &\text{if $p \le t < e$;}\\
    0, &\text{otherwise.}\\
    \end{cases}
\end{equation}

\vspace{-0.5em}{\noindent}For simplicity, we set $l = 0$ and approximate $f_i(t)$, the waveform of attacker $i$, by $\triangle(t; 0, p_i, e_i, m_i)$.
%\begin{equation}
%f_i(t) \simeq \triangle(t; 0, p_i, e_i, m_i)
%\end{equation}
%
%
%We do not assume that $h$ is positive, and this equation can thus represent both positive and negative waveforms without modification.
%
%\footnote{This formula can represent either positive or negative waveforms without modification since we do not assume $m>0$.}
%We assume that each attacker waveform $f_i$ is modeled by a piecewise linear triangular waveform. This is good enough and suitable for fast computation. This is shown in Fig.~\ref{f-triangle}.
%
%
The timing window of attacker $i$ is represented by interval $[a_i,b_i]$, where $a_i$ and $b_i$ indicate the earliest and latest arrival time, respectively.
%, where $a$ and $b$ represent the earliest and latest arrival of the attacker waveform, as shown in Fig.~\ref{f-timing-window}.
%
For convenience, let $\mathbf{\Theta}_i = (p_i, e_i, m_i, a_i, b_i)$ for attacker $i$.

%we collect the parameters used for representing an attacker and its timing window into vector $\mathbf{\Theta}$ given by
%\begin{equation}
%\mathbf{\Theta} = (p, e, m, a, b).
%\end{equation}

%% file: objective.tex
%\begin{equation}
%W^*=
%    \begin{cases}
%    \max_t \left \{ \sum_{i=1}^{n} f_i(t-s_i), 0 \right \} , &\text{if victim is logic 0;}\\
%    \min_t \left \{ \sum_{i=1}^{n} f_i(t-s_i), 0 \right \}, &\text{otherwise.}\\
%    \end{cases}
%\end{equation}

% and denote the waveform of attacker $i$ by function $f_i(t)$.
%
%, where $t$ represents time. %, where the domain and codomain of function $f_i$ are time and voltage, respectively.
%
%

Given a chain of $n$ attackers, let $W$ be the magnitude of the composite waveform of the $n$ attackers and $s_i$ the time amount by which attacker $f_i(t)$ should be shifted to produce $W$. Let $\mathbf{S} = \{s_1, s_2, \ldots, s_n \}$ and assume a function $J: \mathbb{R}^n \rightarrow \mathbb{R}$, where $\mathbb{R}$ denotes the set of all real numbers. FRAME is to determine the worst-case magnitude
\vspace{-0.5em}
\begin{equation}\label{e-obj-w}
W^*=\max_t \left | \sum_{i=1}^{n} f_i(t-s_i) \right | \triangleq J(\mathbf{S})
\end{equation}

\vspace{-0.5em}{\noindent}and an instance of the corresponding attacker alignment
\vspace{-0.25em}
\begin{equation}\label{e-obj-s}
\mathbf{S}^* = \arg\max_{\mathbf{S}} J(\mathbf{S})
\end{equation}

\vspace{-0.5em}{\noindent}under the constraints that $s_i \in [a_i, b_i ]$ for $i=1,2,\ldots,n$. Using $W^*$, we would be able to identify noisy nets that need further analysis. High-accuracy simulators can be invoked with the alignment information $\mathbf{S}^*$.

Eqs.~(\ref{e-obj-w}-\ref{e-obj-s}) might be solved by conventional search algorithms or by a dynamic programming argument if discrete-time waveforms are assumed.
However, such approaches would be overkill given that only an instance of the worst-case solution would suffice for typical static noise analysis. Furthermore, the uniqueness of the solution to the above equations not being guaranteed, such approaches may incur excessive running time with little gain.
FRAME provides a rapid solution to the above equations by following the procedures outlined in Algorithm~\ref{a-frame}. Note that FRAME typically needs to be invoked twice per chain, once for each direction (i.e. rising and falling) of the chain head input, since the attacker parameters change according to this direction.

%The details of computing $W^*$ and $\mathbf{S}^*$ are presented in Sections~\ref{ss-env} and \ref{ss-sigma}, respectively.

%\begin{algorithm}[t]
%\small
%\caption{The FRAME method}
%\label{a-frame}
%\begin{algorithmic}[1]
%\Procedure{Frame}{}
%\State{\texttt{/* Rising input to chain head */}}
%\State Triangularize attackers\Comment{Section~\ref{ss-model-corr}}
%\State Construct envelopes $E$\Comment{Section~\ref{ss-env}}
%\State Derive $\sigma$\ function from $E$ \Comment{Section~\ref{ss-sigma}}
%%\State Represent $E$ and $\sigma$ by vertices\Comment{Section~\ref{ss-add}}
%\State Calculate $W^*_\text{r}$ and $\mathbf{S}^*_\text{r}$ \Comment{Section~\ref{ss-worst}}
%\State{\texttt{/* Falling input to chain head */}}
%\State{Repeat Lines 3-6 to get $W^*_\text{f}$ and $\mathbf{S}^*_\text{f}$}
%\State \textbf{return} $W^*=\max \left \{W^*_\text{r},W^*_\text{f}\right \}$ and matching $\mathbf{S}^*$
%\EndProcedure
%\end{algorithmic}
%\end{algorithm}

\begin{algorithm}[t]
\small
\caption{The FRAME method}
\label{a-frame}
\renewcommand\algorithmicrequire{\textbf{input}:}
\renewcommand\algorithmicensure{\textbf{output}:}
\begin{algorithmic}[1]
\Require ~~$\mathbf{\Theta} = (p, e, m, a, b)$ for each attacker bump\Comment{Section~\ref{ss-model-corr}}
\Ensure $W^*$ and $\mathbf{S}^*$\Comment{Section~\ref{ss-objective}}
\Procedure{Frame}{}
%\State Triangularize attackers\Comment{Section~\ref{ss-model-corr}}
\For{each attacker}
\State Construct envelopes $E$\Comment{Section~\ref{ss-env}}
\State Derive $\sigma$\ function from $E$ \Comment{Section~\ref{ss-sigma}}
\State Represent $E$ and $\sigma$ by vertices\Comment{Section~\ref{ss-add}}
\EndFor
\State Calculate $W^*$ and $\mathbf{S}^*$ \Comment{Section~\ref{ss-worst}}
%\State \textbf{return} $W^*$ and $\mathbf{S}^*$
\EndProcedure
\end{algorithmic}
\end{algorithm}

%% file: envelope.tex
We utilize the notion of \emph{envelopes} to determine efficiently the magnitude of a worst-case composite waveform. Informally, an envelope is a waveform that represents the maximum or minimum voltage observable at each time point, as an attacker waveform is slid over its timing window. The idea of using an envelope has some resemblance to the work by~\cite{tehrani00}, but the envelope we propose is more general and powerful in that both maximum and minimum values can be tracked and that alignment information is accompanied.

\begin{figure}
\centering
\subfigure[Probing voltage at $t=t_x$] {
    \psfrag{t}{$t$}
    \psfrag{0}{0}
    \psfrag{1}{1}
    \psfrag{2}{2}
    \psfrag{3}{3}
    \psfrag{4}{4}
    \psfrag{5}{5}
    \psfrag{6}{6}
    \psfrag{7}{7}
    \psfrag{8}{8}
    \psfrag{a}{$a$}
    \psfrag{b}{$b$}
    \psfrag{c}{$a+e$}
    \psfrag{d}{~~$b+e$}
    \psfrag{w}{timing window}
    \psfrag{x}{$t_x$}
    \psfrag{p}{\!$a+p$}
    \includegraphics[height=0.75in, width=0.9\linewidth]{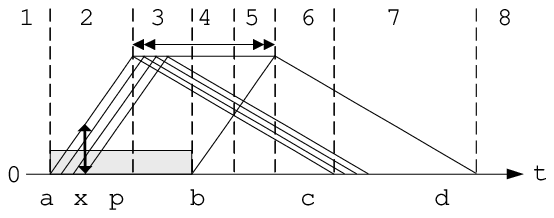}
    \label{f-env-intuition-a}
}
\subfigure[Computing max/min at $t_x$ depends on time zone] {
    \renewcommand{\arraystretch}{1.1}
%    \caption{The min and max values observable at $t$.}
    \begin{tabular}{|c||c|c|c|c|c|c|}\hline
    Zone &  2 & 3 & 4& 5 & 6& 7\\\hline\hline
    Max & $\frac{m}{p}(t-a)$ & \multicolumn{3}{c|}{$m$} & \multicolumn{2}{c|}{$\frac{-m}{e-p}(t-e-b)$} \\[0.4ex]\hline
    Min & \multicolumn{2}{c|}{$0$} & $\frac{m}{p}(t-b)$ & \multicolumn{2}{c|} {$\frac{-m}{e-p}(t-a-e)$} & \multicolumn{1}{c|}{$0$}\\[0.4ex]\hline
    %min ($b \ge a+ e$) & \multicolumn{8}{c|}{0}\\\hline
    \end{tabular}
    \label{f-env-intuition-b}
}
\caption{Concept of envelopes.}
\label{f-env-intuition}
\end{figure}

For a more precise definition, consider Fig.~\ref{f-env-intuition-a} that shows the trajectory of an attacker waveform over its timing window.
Assume that $f(t) =\triangle(t; 0, p, e, m)$ and that $f$ can be shifted by $s \in [a,b]$. For the sake of explanation, also assume that $b < a+ e$.
Consider the probe point at time $t=t_x$ in Fig.~\ref{f-env-intuition-a}. The vertical arrow then indicates the range of voltage values that can be measured at $t_x$, as $s$ varies from $a$ to $b$.
The maximum voltage observable at $t_x$ is $\frac{m}{p}(t_x - a)$, which happens when $s = a$. Similarly, the minimum voltage should be 0, which is achieved when $t_x \le s \le b$. (Recall that we are varying $s$ over fixed $t_x$.)

Above we assumed that $t_x$ is in zone 2 or $a \le t_x \le b$; by varying $t_x$ over the entire time horizon and recording the maximum and minimum at every $t_x$, we obtain envelopes. As summarized in the table in Fig.~\ref{f-env-intuition-b}, the formula to compute the minimum and maximum values observable at $t_x$ varies depending upon where $t_x$ is placed. The second line of the table lists the formula to get the maximum for each time zone. For the minimum, the relative locations of $b$ and $a+e$ also matters. The third line shows only the formula to calculate the minimum for $b < a + e$; otherwise, the minimum is always zero. The entries for zones 1 and 8 are not shown in the table since the values are all zero.

%%%%%%%%%%%%%%%%%%%%%%%%%%%%%%%%%%%%%%%%%%%%%%%%%%%%%%%%%%%%%%%%%%%%%%%%%%
%%%%%%%%%%%%%%%%%%%%%%%%%%%%%%%%%%%%%%%%%%%%%%%%%%%%%%%%%%%%%%%%%%%%%%%%%%
%% ENVELOPE (Inituitive explanation)
%%%%%%%%%%%%%%%%%%%%%%%%%%%%%%%%%%%%%%%%%%%%%%%%%%%%%%%%%%%%%%%%%%%%%%%%%%
%%%%%%%%%%%%%%%%%%%%%%%%%%%%%%%%%%%%%%%%%%%%%%%%%%%%%%%%%%%%%%%%%%%%%%%%%%

%%%%%%%%%%%%%%%%%%%%%%%%%%%%%%%%%%%%%%%%%%%%%%%%%%%%%%%%%%%%%%%%%%%%%%%%%%
%%%%%%%%%%%%%%%%%%%%%%%%%%%%%%%%%%%%%%%%%%%%%%%%%%%%%%%%%%%%%%%%%%%%%%%%%%

Now we formally define two types of envelopes --- \emph{outer} and \emph{inner} envelopes. The outer envelope $E_{\text{outer}}$ of attacker $f(t)$ represented by $\triangle(t; 0, p, e, m)$ is a function of time $t$ parameterized by $\mathbf{\Theta} = (p, e, m, a, b)$:
\vspace{-1em}
\begin{multline*}
E_{\text{outer}}\left (t; \mathbf{\Theta} \right ) =
%E_{\text{outer}}\left (t; (p, e, m, a, b) \right ) = \\
\begin{cases}
    \frac{m}{p}(t-a), &\text{if $a \le t < a+ p$;}\\
    m, &\text{if $a + p \le t < b + p$;}\\
    \frac{-m}{e-p}(t-e-b), &\text{if $b + p \le t < b + e$;}\\
    0, &\text{otherwise.}\\
    \end{cases}
\end{multline*}

\vspace{-0.5em}{\noindent}For a positive waveform ($m>0$), the outer envelope is the trace of the maximum, as shown in Fig.~\ref{f-env-outer-pos}. For a negative waveform ($m<0$), the outer envelope is the trace of the minimum, as shown in Fig.~\ref{f-env-outer-neg}.
The inner envelope of an attacker can have one of the two shapes, depending upon the parameters and timing window used:
\vspace{-1em}
\begin{multline*}
E_{\text{inner}}\left (t; \mathbf{\Theta} \right ) =
%E_{\text{inner}}\left ( t; (p, e, m, a, b) \right ) = \\
    \begin{cases}
%    \triangle\left(t;b, b+p\left(\frac{a+e-b}{e}\right), a+e, m\left(\frac{a+e-b}{e}\right)\right), &\text{if $b < a+ e$;}\\
    \triangle\left(t;b, b+p\delta, a+e, m\delta\right), &\text{if $b < a+ e$;}\\
    0, &\text{otherwise.}\\
    \end{cases}
\end{multline*}

\vspace{-0.5em}{\noindent}where $\delta \triangleq \frac{a+e-b}{e}$.
The inner envelope of a positive waveform traces the minimum of the waveform. If $b<a+e$, the inner envelope has a triangular shape as shown in Fig.~\ref{f-env-inner-pos}; otherwise, the inner envelope is always zero as shown in Fig.~\ref{f-env-inner-pos-null}. The inner envelope of a negative waveform traces the maximum of the waveform as shown in Fig.~\ref{f-env-inner-neg} and~\ref{f-env-inner-neg-null}. Table~\ref{t-env-meaning} summarizes the meaning of outer and inner envelopes for positive and negative waveforms. %Fig.~\ref{f-env} shows all possible envelopes, assuming that attacker $f(t)$ is represented by $\triangle(t; 0, p, e, m)$.

\begin{table}
\caption{The meaning of outer/inner envelopes.}
\label{t-env-meaning}
    \begin{tabular}{|c||c|c|}\hline
    Wave &  Outer envelope &Inner envelope \\\hline\hline
    Positive & max trace: Fig.~\ref{f-env-outer-pos} & min trace: Fig.~\ref{f-env}(b),(c)\\\hline
    Negative & min trace: Fig.~\ref{f-env-outer-neg} & max trace: Fig.~\ref{f-env-neg}(b),(c)\\\hline
    \end{tabular}
\end{table}

%% file: sigma.tex
The FRAME method provides not only the magnitude of a worst-case combined attacker waveform but also the corresponding attacker alignment. Using this alignment information, high-accuracy simulators can be invoked for further analysis. Each FRAME envelope is accompanied by the time-shift information named the \emph{$\sigma$ function}. Recall in Fig.~\ref{f-env-intuition-a} the maximum voltage at $t_x$ is achieved when attacker $f$ is shifted by $s=a$. Informally, the $\sigma$ function is to record this shift information for every point on an envelope.

The shift amount $s$ that makes the voltage at $t_x$ maximum or minimum is sometimes not unique. For instance, the minimum at $t_x$ can be achieved as long as $t_x \le s \le b$ in Fig.~\ref{f-env-intuition-a}. This fact can be represented more formally by $R_\sigma$, a binary relation from $\mathbb{R}$ to $[a,b]$ such that
$R_\sigma = \{ (t, s) | f(t-s) = E(t) \}$
%\begin{equation}
%R_\sigma = \{ (t, s) | f(t-s) = E(t) \}
%\end{equation}
for $t\in \mathbb{R}$ and $s \in [a,b]$. That is, for any given $t$, $R_\sigma$ provides the set of all values by which an attacker should be shifted to have the maximum or minimum value recorded on its envelope. The gray areas and nonzero lines in each subplot of Fig.~\ref{f-sigma} represent the $R_\sigma$ relation for each of the envelopes shown in Fig.~\ref{f-env} and~\ref{f-env-neg}.
%

%\input{f-sigma}

%Since we do not all instances of the worst-case alignment is ,

Since we aim at finding some (not all) instances of the worst-case alignment, we do not use $R_\sigma$ as it is but derive a function $\sigma: \mathbb{R} \rightarrow [a,b]$ from the $R_\sigma$ relation. It is this $\sigma(t)$ function that is used in the downstream procedures. The thick bold line segments drawn in Fig.~\ref{f-sigma-outer} represent our choice for $\sigma(t)$ for the outer envelopes presented in Fig.~\ref{f-env-outer-pos} and~\ref{f-env-outer-neg}. Note that this arbitrary choice does not compromise the correctness of the solution we find; introducing $\sigma$ functions helps enhance computational efficiency by limiting the search space. Fewer solutions would be found by adopting $\sigma$, but the number of solutions is not important in our context. The $\sigma$ function we adopt for the inner envelopes presented in Fig.~\ref{f-env-inner-pos} and~\ref{f-env-inner-neg} is indicated by dark lines in Fig.~\ref{f-sigma-inner}. Finally, the thick bold lines in Fig.~\ref{f-sigma-inner-null} represent the $\delta$ function of our choice for the inner envelopes in Fig.~\ref{f-env-inner-pos-null} and~\ref{f-env-inner-neg-null}. Table~\ref{t-sigma} summarizes the formulae for the $\sigma$ functions we use.

%The $\sigma$ functions we use are summarized in Table~\ref{t-sigma}, where constant $\delta \triangleq \frac{a+e-b}{e}$.

\begin{table}
\centering
\caption{The $\sigma$ functions we use ($\delta \triangleq \frac{a+e-b}{e}$).}
\label{t-sigma}
\begin{tabular}{|c||c|c|}\hline
\parbox[b][1.1em]{7em}{For $E_{\text{outer}}(t)$:} &
\multicolumn{2}{c|}{\parbox{7em}{For $E_{\text{inner}}(t)$:}}
\\\hline\hline
\parbox[c][6.7em]{8em}{
\vspace{-0.25cm}
    \begin{multline*}
    \sigma_{\text{}}(t;\mathbf{\Theta}) = \\
        \begin{cases}
            a, &\text{if $t < a$;}\\
            t, &\text{if $a \le t < b$;}\\
            b, &\text{otherwise.}\\
        \end{cases}
    \end{multline*}
} &
\parbox[c]{8em}{
\vspace{-0.25cm}
    \begin{multline*}
    \sigma_{\text{}}(t;\mathbf{\Theta}) = \\
    \begin{cases}
        b, &\text{if $t < b + p\delta$;}\\
        a, &\text{otherwise.}\\
    \end{cases}
\end{multline*}
%(when $b < a+ e$)
} &
\parbox[c]{7em}{
\vspace{-0.25cm}
    \begin{multline*}
    \sigma_{\text{}}(t;\mathbf{\Theta}) = \\
    \begin{cases}
        b, &\text{if $t < b$;}\\
        a, &\text{otherwise.}\\
    \end{cases}
\end{multline*}
%    (when $b \ge a+ e$)
}\\\hline
See Fig.~\ref{f-sigma-outer}&\ref{f-sigma-inner}: $b < a+ e$ & \ref{f-sigma-inner-null}: $b \ge a+e$\\\hline
\end{tabular}
\end{table}

%\begin{equation}
%\sigma_{\text{outer}}(t;\mathbf{\Theta})=
%\begin{cases}
%    a, &\text{if $t < a$;}\\
%    t, &\text{if $a \le t < b$;}\\
%    b, &\text{otherwise.}\\
%\end{cases}
%\end{equation}
%which is indicated in Fig.~\ref{f-sigma-outer} by dark line segments. The $\sigma$ function for the inner envelopes shown in Fig.~\ref{f-env-inner-pos} and~\ref{f-env-inner-neg} (for $b < a+ e$) is \begin{equation}
%\sigma_{\text{inner}}(t;\mathbf{\Theta})=
%\begin{cases}
%    b, &\text{if $t < b + p\delta$;}\\
%    a, &\text{otherwise.}\\
%    \end{cases}
%\end{equation}
%where $\delta \triangleq \frac{a+e-b}{e}$, as marked by dark lines in Fig.~\ref{f-sigma-inner}. Finally, the $\sigma$ function for the remaining case (inner, $b \ge a+ e$) is given by
%\begin{equation}
%\sigma_{\text{inner}}(t;\mathbf{\Theta})=
%\begin{cases}
%    b, &\text{if $t < b$;}\\
%    a, &\text{otherwise.}\\
%    \end{cases}
%\end{equation}
%as drawn in Fig.~\ref{f-sigma-inner-null}.

%\begin{table}
%\centering
%\caption{Function $\sigma(t)$}
%\label{t-sigma}
%\begin{tabular}{|rl|rl|rl|}\hline
%\multicolumn{2}{|c|}{$\sigma_{\text{outer}}(t)$} & \multicolumn{2}{c|}{$\sigma_{\text{outer}}(t)$} & \multicolumn{2}{c|}{$\sigma_{\text{outer}}(t)$}\\\hline\hline
%$a$, & if $t < a$; & $b$, & if $t < b+p\delta$; & $b$, & if $t < b$; \\
%$t$, & if $a \le t < b$; & & & & \\
%$b$, & otherwise. & $a$, & otherwise. & $a$, & otherwise.\\\hline
%\end{tabular}
%\end{table}

%\left\{
%  \begin{array}{ll}
%    , & \hbox{;} \\
%    , & \hbox{.}
%  \end{array}
%\right.

%% file: manipulation.tex
The piecewise linear assumption enables us to represent an envelope only by the vertices joining line segments. For instance, the outer envelope drawn in Fig.~\ref{f-env-outer-pos} can be represented by using the four vertices shown in Fig.~\ref{f-vertex-outer-pos}. In this manner, we can avoid handling an envelope in its continuous form; any intermediate values can be deduced from the vertices used. Similarly, we can save the need to manipulate the whole $\sigma$ function by annotating each vertex with its $\sigma$ value.
Each vertex in our representation thus has four attributes denoted by tuple $(t, V, [s_L, s_R])$, where $t$ and $V$ represent the time and voltage associated with the vertex; $s_L$ and $s_R$ are the left and right limits of $\sigma$ at $t$:
$s_L = \lim_{\epsilon \rightarrow 0^-} \sigma(t+\epsilon)$ and
$s_R = \lim_{\epsilon \rightarrow 0^+} \sigma(t+\epsilon)$.
%\begin{equation}
%s_L = \lim_{\epsilon \rightarrow 0^-} \sigma(t+\epsilon)~\text{and~}
%s_R = \lim_{\epsilon \rightarrow 0^+} \sigma(t+\epsilon).
%\end{equation}
%
Storing both limits is to handle the discontinuities appearing in the $\sigma$ functions for inner envelopes. FRAME represents each attacker waveform and its timing window compactly by using these annotated vertices. Fig.~\ref{f-vertex} shows the representations of the outer and inner envelopes presented in Fig.~\ref{f-env}. The representations of the envelopes in Fig.~\ref{f-env-neg} are similarly defined and not shown.

%\begin{eqnarray}
%s_L &=& \lim_{\epsilon \rightarrow 0} s(t-\epsilon)\\
%s_R &=& \lim_{\epsilon \rightarrow 0} s(t+\epsilon)
%\end{eqnarray}
%left and right shift amount defined as follows

%In order to restore the original envelope curve, each vertex is annotated with the time and voltage values it has in the original curve. In addition, we also annotate with each vertex how much it

% Associated with each vertex $v$ is tuple $(t, V, [s_L, s_R])$, where $t$ and $V$ are the time and voltage values of $v$, respectively. The  $s_L$ and $s_R$ are the time shift amount of the left and right to the vertex $v$, respectively.

%\subsubsection{Envelope addition}

%\input{f-representation}

%% file: worst.tex
The last step of FRAME is to manipulate envelopes and $\sigma$ functions to determine the worst-case composite waveform and attacker alignment. Since the envelope of an attacker waveform represents the maximum or minimum voltage for each time point, adding the envelopes of multiple attackers reveals the maximum or minimum voltage of the combined waveform of these attackers. More precisely, the worst-case magnitude $W^*$ of a combined waveform is given by
\vspace{-0.5em}
\begin{equation}
W^*  = \max_{t} \left \{ \left | E_{\text{max}}(t) \right |, \left | E_{\text{min}}(t) \right | \right \}
\end{equation}

\vspace{-0.5em}{\noindent}where $E_{\text{max}}(t)$ and $E_{\text{min}}(t)$ represent the envelopes to find the maximum and minimum voltage, respectively. For a positive (negative) attacker, its outer (inner) envelope represents the maximum value. Thus, $E_{\text{max}}(t)$ is simply the sum of all positive outer and negative inner envelopes. Similarly, $E_{\text{min}}(t)$ can be calculated by adding all positive inner and negative outer envelopes. % Thus, $E_{\text{max}}(t)$ and $E_{\text{min}}(t)$ are given by
More precisely,
\vspace{-0.5em}
\begin{eqnarray}
E_{\text{max}}(t) &\!\!\!=\!\!\!& \sum_{i\in \mathbf{I}^+} E_{\text{outer}}(t;\mathbf{\Theta}_i) + \sum_{i\in \mathbf{I}^-} E_{\text{inner}}(t;\mathbf{\Theta}_i)\\
E_{\text{min}}(t) &\!\!\!=\!\!\!& \sum_{i\in \mathbf{I}^+} E_{\text{inner}}(t;\mathbf{\Theta}_i) + \sum_{i\in \mathbf{I}^-} E_{\text{outer}}(t;\mathbf{\Theta}_i)
\end{eqnarray}

\vspace{-0.5em}{\noindent}where $\mathbf{I}^+ = \{ i | m_i \ge 0 \}$, $\mathbf{I}^- = \{ i | m_i < 0 \}$
%\begin{eqnarray}
%\mathbf{I}^+ &=& \{ i |~m_i \ge 0 \}\\
%\mathbf{I}^- &=& \{ i |~m_i < 0 \}
%\end{eqnarray}
and
$|\mathbf{I}^+| + |\mathbf{I}^-| = n$. %That is, $E_{\text{max}}(t)$ should be the sum of the outer envelopes of positive attackers and the inner envelopes of negative attackers since... and $E_{\text{min}}(t)$ is calculated by adding positive inner and negative outer envelopes.
The worst-case alignment $s_i^*$ of attacker $i$ is given by
\vspace{-0.5em}
%\begin{equation}
%%s_i^* = \sigma_i \left (\arg\max_{t} \left \{ \left | E_{\text{max}}(t) \right |, \left | E_{\text{min}}(t) \right | \right \} \right)
%s_i^* = \sigma_i \left (t^* \right )
%\end{equation}
%\begin{eqnarray}
\begin{gather}
s_i^* = \sigma_i \left (t^* \right )\text{, where}\\
t^* = \arg\max_{t} \left \{ \left | E_{\text{max}}(t) \right |, \left | E_{\text{min}}(t) \right | \right \}
\end{gather}
%\end{eqnarray}

\vspace{-0.5em}{\noindent}for $i=1,2,\ldots,n$.
%When multiple envelopes are added, their $\sigma$ functions are not added but of the added envelopes so that the alignment information of each attacker can be retrieved later.
%
%The shift information needed for the worst-case attacker alignment can be retrieved from the $s_L$ and $s_R$ attributes of the vertex representation.
%
Using the representation described in Section~\ref{ss-add}, we can add envelopes simply by manipulating the attributes of the vertices used in the representation. We use the following example for further explanation.

%add two envelopes to compute $|E_{\text{max}}(t)|$.%; $|E_{\text{min}}(t)|$ can be determined similarly.

%Algorithm~1 presented in Fig.~\ref{} outlines the algorithm to add two envelopes. Extending this algorithm to more than two envelopes is straightforward.

%\begin{algorithm}[t]
%%\small
%\caption{Adding two envelopes}
%\label{a-env-add}
%\begin{algorithmic}[1]
%\Function{AddTwoEnvelopes}{$T_1,T_2$}
%\State $E \gets \{\}$
%\ForAll{$t \in (T_1 \cup T_2)$}
%    \If{$t \in (T_1 \cap T_2)$}\Comment{Common vertex}
%        \State create $v_{\text{new}}$
%        \State $v_{\text{new}}.t \gets v.t$
%        \State $v_{\text{new}}.V \gets v.V$
%    \ElsIf{}
%    \Else
%    \EndIf
%\EndFor
%\State \textbf{return} $E$
%\EndFunction
%\end{algorithmic}
%\end{algorithm}

\input{f-envelope}
\input{f-sigma}
\input{f-representation}
\input{f-example}

\vspace{-0.5em}
\begin{example}
\emph{
\textbf{Fig.~\ref{f-example}(a)} shows two attacker waveforms in the left column. They are converted to the vertex-based envelope representations in the right column and then added to $E_{\text{max}}(t)$. \textbf{Fig.~\ref{f-example}(b)} lists the attribute of each vertex before addition. An empty vertex attribute tuple of an attacker means that this vertex does not appear in the representation of the attacker. Before addition, these empty attributes get filled with values linearly interpolated using existing vertex attributes. For instance, \textbf{Fig.~\ref{f-example}(c)} shows that the $v_2$ voltage of attacker 1 is set to the average voltage of $v_1$ and $v_3$ marked by hexagons; the $s_L$ and $s_R$ of the new $v_4$ vertex is set to the average of $s_R$ of $v_3$ and $s_L$ of $v_5$ marked by a rectangle and an ellipse, respectively. Interpolated entries are marked with asterisks. The `Sum' column shows the result of addition. The voltage of a vertex in the result is the sum of the voltage attributes of the vertices of attackers 1 and 2. \textbf{Fig.~\ref{f-example}(d)} shows $E_{\text{min}}(t)$ obtained by switching the roles of outer and inner envelopes. Since $|E_{\text{min}}| > |E_{\text{max}}|$, $W^* = |E_{\text{min}}| =3$. Note that there are three instances of $t^*$. \textbf{Fig.~\ref{f-example}(e)} presents the worst-case waveform $f_1(t-s_1^*) + f_2(t-s_2^*)$, where $s_1^* = \sigma_1(t^*=5)=1$ and $s_2^* = \sigma_2(t^*=5)=2$. The other cases for $t^*=3$ and $4$ are not shown due to the space limit.}\hfill$\Box$
 %That is, $|E_{\text{max}}(t)| = 1(V)$, which occurs at $t=2,3$ or $5$. For $t=2$, we read the line for $v_2$ to discover that attackers 1 and 2 should be shifted by 1 and 2, respectively, to produce $|E_{\text{max}}(t)|$.
\end{example}
%That is, if the worst-case voltage is measured at time $t^*$, examining the tuple chain annotating the vertex at $t^*$ gives the alignment information for each attacker. (If $s_L$ and $s_R$ values are different, we can choose either one.)

%This algorithm can easily be extended to add multiple envelopes
\vspace{-0.75em}
The addition algorithm used in the above example can easily be extended to adding multiple envelopes: Given $n$ envelopes, we first sort the vertices of all envelopes with respect to $t$ and merge them into a single master vertex list. We annotate each vertex in the list with a chain of $n$ attribute tuples, one for each attacker envelope. For each attacker, we examine its chain of attribute tuples and interpolate any empty entries. Finally, for each vertex in the master list, we compute its voltage attribute by accumulating the voltage over the chain of attribute tuples. The worst-case complexity of the algorithm to add multiple envelopes is $O(n\text{log}n + 4n\cdot n + 4n \cdot n) = O(n^2)$.%, $n$ being the number of attackers.

%\subsection{Proposed Algorithm}
%\begin{figure}
%\centering
%\includegraphics[width=\linewidth,height=2in]{box}
%\caption{Proposed algorithm.}
%\label{f-algorithm}
%\end{figure}

%% file: f-envelope.tex
%%%%%%%%%%%%%%%%%%%%%%%%%%%%%%%%%%%%%%%%%%%%%%%%%%%%%%%%%%%%%%%%%%%%%%%%%%
%%%%%%%%%%%%%%%%%%%%%%%%%%%%%%%%%%%%%%%%%%%%%%%%%%%%%%%%%%%%%%%%%%%%%%%%%%
%% ENVELOPE (Positive waveform)
%%%%%%%%%%%%%%%%%%%%%%%%%%%%%%%%%%%%%%%%%%%%%%%%%%%%%%%%%%%%%%%%%%%%%%%%%%
%%%%%%%%%%%%%%%%%%%%%%%%%%%%%%%%%%%%%%%%%%%%%%%%%%%%%%%%%%%%%%%%%%%%%%%%%%
\begin{figure*}
\centering
\psfrag{t}{$t$}
\psfrag{0}{$0$}
\psfrag{V}{$V$}
\subfigure[Outer] {
    \psfrag{a}{$\!\!a$}
    \psfrag{b}{$\!\!b+e$}
    \psfrag{x}{$\frac{m}{p}(t-a)$}
    \psfrag{y}{$\!\!\!\!\!\!\!\!\frac{-m}{e-p}(t-e-b)$}
    \psfrag{h}{$m$}
    \psfrag{p}{$\!\!a+p$}
    \psfrag{q}{$\!\!\!\!b+p$}
    \includegraphics[width=0.31\linewidth]{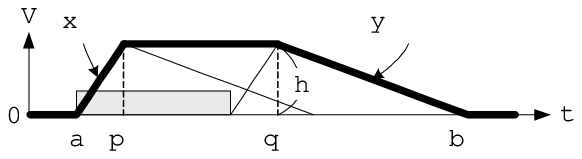}
    \label{f-env-outer-pos}
}
\subfigure[Inner ($b < a+e$)] {
    \psfrag{a}{$\!\!a$}
    \psfrag{b}{$\!\!b+e$}
    \psfrag{p}{$b$}
    \psfrag{q}{$\!\!\!a+e$}
%    \psfrag{z}{$\!\!\!\!m\left ( \frac{a+e-b}{e}\right )$}
%    \psfrag{r}{$b+p\left ( \frac{a+e-b}{e}\right )$}
%    \psfrag{x}{$\!\!\!\!\triangle \left(t; b, b+p \left ( \frac{a+e-b}{e} \right ), a + e, m \left ( \frac{a+e-b}{e} \right ) \right)$}
    \psfrag{x}{\!\!\!\!\!$\quad\triangle \left(t; b, b+p \delta, a + e, m \delta \right)$}
%    \psfrag{g}{\makebox{$b < a+e$}}
%    \psfrag{h}{otherwise}
    \includegraphics[width=0.31\linewidth]{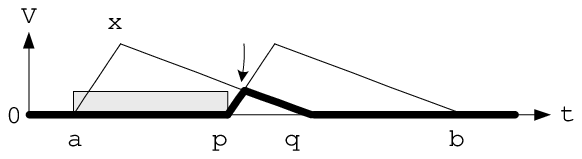}
    \label{f-env-inner-pos}
}
\subfigure[Inner ($b \ge a+e$)] {
    \psfrag{a}{$\!\!a$}
    \psfrag{b}{$\!\!b+e$}
    \psfrag{q}{$b$}
    \psfrag{p}{$\!\!a+e$}
    \includegraphics[width=0.31\linewidth]{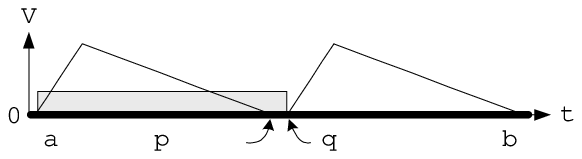}
    \label{f-env-inner-pos-null}
}
\caption{Outer and inner envelopes for a \emph{positive} attacker waveform ($\delta \triangleq \frac{a+e-b}{e}$).}
\label{f-env}
\end{figure*}
%%%%%%%%%%%%%%%%%%%%%%%%%%%%%%%%%%%%%%%%%%%%%%%%%%%%%%%%%%%%%%%%%%%%%%%%%%
%%%%%%%%%%%%%%%%%%%%%%%%%%%%%%%%%%%%%%%%%%%%%%%%%%%%%%%%%%%%%%%%%%%%%%%%%%

%%%%%%%%%%%%%%%%%%%%%%%%%%%%%%%%%%%%%%%%%%%%%%%%%%%%%%%%%%%%%%%%%%%%%%%%%%
%%%%%%%%%%%%%%%%%%%%%%%%%%%%%%%%%%%%%%%%%%%%%%%%%%%%%%%%%%%%%%%%%%%%%%%%%%
%% ENVELOPE (Negative waveform)
%%%%%%%%%%%%%%%%%%%%%%%%%%%%%%%%%%%%%%%%%%%%%%%%%%%%%%%%%%%%%%%%%%%%%%%%%%
%%%%%%%%%%%%%%%%%%%%%%%%%%%%%%%%%%%%%%%%%%%%%%%%%%%%%%%%%%%%%%%%%%%%%%%%%%
\begin{figure*}
\centering
\psfrag{t}{$t$}
\psfrag{0}{$0$}
\psfrag{V}{$V$}

\subfigure[Outer] {
    \psfrag{a}{$\!\!a$}
    \psfrag{b}{$\!\!b+e$}
    \psfrag{x}{$\frac{m}{p}(t-a)$}
    \psfrag{y}{$\frac{-m}{e-p}(t-e-b)$}
    \psfrag{h}{$m$}
    \psfrag{p}{$\!\!a+p$}
    \psfrag{q}{$\!\!\!\!b+p$}
    \includegraphics[width=0.31\linewidth]{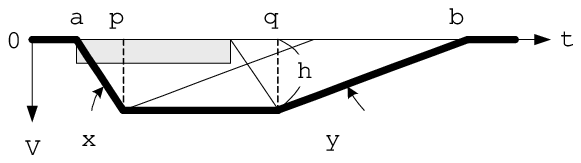}
    \label{f-env-outer-neg}
}
\subfigure[Inner ($b < a+e$)] {
    \psfrag{a}{$\!\!a$}
    \psfrag{b}{$\!\!b+e$}
    \psfrag{p}{$b$}
    \psfrag{q}{$\!\!\!a+e$}
%    \psfrag{z}{$\!\!\!\!m\left ( \frac{a+e-b}{e}\right )$}
%    \psfrag{r}{$b+p\left ( \frac{a+e-b}{e}\right )$}
%    \psfrag{x}{$\!\!\!\!\triangle \left(t; b, b+p \left ( \frac{a+e-b}{e} \right ), a + e, m \left ( \frac{a+e-b}{e} \right ) \right)$}
\psfrag{x}{$\quad\triangle \left(t; b, b+p\delta, a + e, m\delta \right)$}
    \psfrag{g}{\makebox{$b < a+e$}}
    \psfrag{h}{otherwise}
    \includegraphics[width=0.31\linewidth]{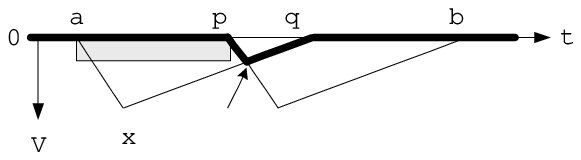}
    \label{f-env-inner-neg}
}
\subfigure[Inner ($b \ge a+e$)] {
    \psfrag{a}{$\!\!a$}
    \psfrag{b}{$\!\!b+e$}
    \psfrag{q}{$b$}
    \psfrag{p}{$\!\!a+e$}
    \includegraphics[width=0.31\linewidth]{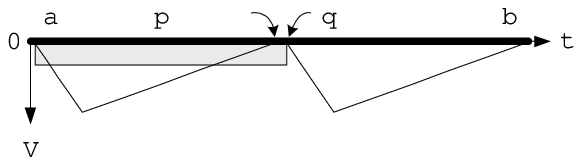}
    \label{f-env-inner-neg-null}
}
\caption{Outer and inner envelopes for a \emph{negative} attacker waveform ($\delta \triangleq \frac{a+e-b}{e}$).}
\label{f-env-neg}
\end{figure*}
%%%%%%%%%%%%%%%%%%%%%%%%%%%%%%%%%%%%%%%%%%%%%%%%%%%%%%%%%%%%%%%%%%%%%%%%%%
%%%%%%%%%%%%%%%%%%%%%%%%%%%%%%%%%%%%%%%%%%%%%%%%%%%%%%%%%%%%%%%%%%%%%%%%%%

%% file: f-sigma.tex
%%%%%%%%%%%%%%%%%%%%%%%%%%%%%%%%%%%%%%%%%%%%%%%%%%%%%%%%%%%%%%%%%%%%%%%%%%
%%%%%%%%%%%%%%%%%%%%%%%%%%%%%%%%%%%%%%%%%%%%%%%%%%%%%%%%%%%%%%%%%%%%%%%%%%
%% SIGMA
%%%%%%%%%%%%%%%%%%%%%%%%%%%%%%%%%%%%%%%%%%%%%%%%%%%%%%%%%%%%%%%%%%%%%%%%%%
%%%%%%%%%%%%%%%%%%%%%%%%%%%%%%%%%%%%%%%%%%%%%%%%%%%%%%%%%%%%%%%%%%%%%%%%%%
\begin{figure*}
\psfrag{t}{$t$}
\psfrag{s}{$s$}
\psfrag{0}{$0$}
\centering
\mbox{
\subfigure[For outer envelopes] {
    \psfrag{a}{$a$}
    \psfrag{b}{$b$}
    \psfrag{c}{$\!\!b+e$}
    \includegraphics[width=0.31\linewidth]{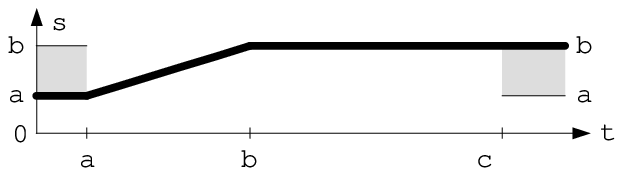}
    \label{f-sigma-outer}
}
\subfigure[For inner envelopes ($b < a+ e$)] {
    \psfrag{a}{$a$}
    \psfrag{b}{$b$}
    \psfrag{c}{$\!\!\!b+e$}
    \psfrag{d}{$\!\!\!\!\!\!a+e$}
    \psfrag{e}{$b+p\delta$}
%    \psfrag{e}{$b+p\left ( \frac{a+e-b}{e} \right )$}
    \includegraphics[width=0.31\linewidth]{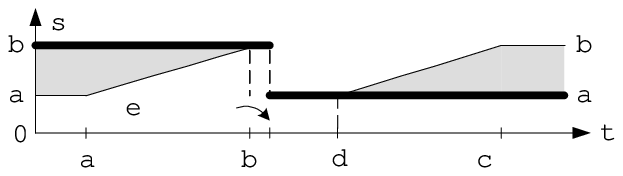}
    \label{f-sigma-inner}
}
\subfigure[For inner envelopes ($b \ge a+ e$)] {
    \psfrag{a}{$a$}
    \psfrag{b}{$b$}
    \psfrag{c}{$\!\!\!b+e$}
    \psfrag{d}{$\!\!a+e$}
    \includegraphics[width=0.31\linewidth]{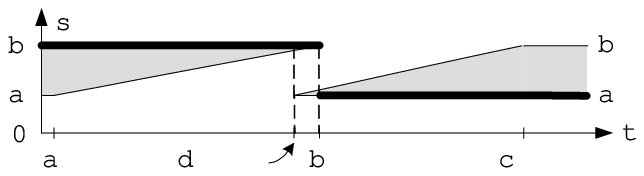}
    \label{f-sigma-inner-null}
}
}
\caption{Relation $R_\sigma$ and function $\sigma(t)$ for the envelopes in Fig.~\ref{f-env} and Fig.~\ref{f-env-neg}.}
\label{f-sigma}
\end{figure*}
%%%%%%%%%%%%%%%%%%%%%%%%%%%%%%%%%%%%%%%%%%%%%%%%%%%%%%%%%%%%%%%%%%%%%%%%%%
%%%%%%%%%%%%%%%%%%%%%%%%%%%%%%%%%%%%%%%%%%%%%%%%%%%%%%%%%%%%%%%%%%%%%%%%%%

%% file: f-representation.tex
%%%%%%%%%%%%%%%%%%%%%%%%%%%%%%%%%%%%%%%%%%%%%%%%%%%%%%%%%%%%%%%%%%%%%%%%%%
%%%%%%%%%%%%%%%%%%%%%%%%%%%%%%%%%%%%%%%%%%%%%%%%%%%%%%%%%%%%%%%%%%%%%%%%%%
%% REPRESENTATION
%%%%%%%%%%%%%%%%%%%%%%%%%%%%%%%%%%%%%%%%%%%%%%%%%%%%%%%%%%%%%%%%%%%%%%%%%%
%%%%%%%%%%%%%%%%%%%%%%%%%%%%%%%%%%%%%%%%%%%%%%%%%%%%%%%%%%%%%%%%%%%%%%%%%%
\begin{figure*}
\centering
\psfrag{t}{$t$}
\psfrag{0}{$0$}
\psfrag{V}{$V$}
\subfigure[Outer] {
    \psfrag{x}{$(a+p,m,[a,a])$}
    \psfrag{z}{\qquad$(b+p,m,[b,b])$}
    \psfrag{a}{\!\!$a$}
    \psfrag{b}{$b$}
    \psfrag{c}{$(a,0,[a,a])$}
    \psfrag{d}{\quad$(b+e,0,[b,b])$}
    \includegraphics[width=0.31\linewidth]{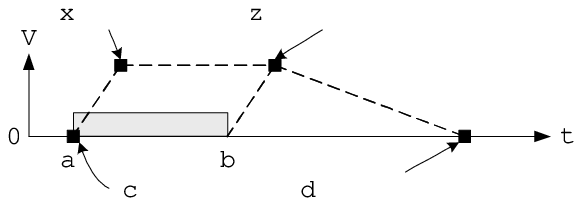}
    \label{f-vertex-outer-pos}
}
\subfigure[Inner ($b < a+ e$)] {
%    \psfrag{x}{$\left(b+p\left (\frac{a+e-b}{e}\right ),m\left (\frac{a+e-b}{e}\right ) ,[b,a] \right )$}
    \psfrag{x}{$\left(b+p\delta,m\delta ,[b,a] \right )$}
    \psfrag{a}{$a$}
    \psfrag{b}{$b$}
    \psfrag{c}{$(b,0,[b,b])$}
    \psfrag{d}{\quad$(a+e,0,[a,a])$}
    \includegraphics[width=0.31\linewidth]{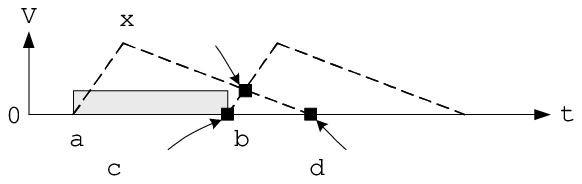}
    \label{f-vertex-inner-pos}
}
\subfigure[Inner ($b \ge a+ e$)] {
    \psfrag{x}{$(b,0,[b,a])$}
    \psfrag{a}{\!\!$a$}
    \psfrag{b}{$b$}
    \psfrag{c}{\!\!\!\!\!$a+e$}
    \includegraphics[width=0.31\linewidth]{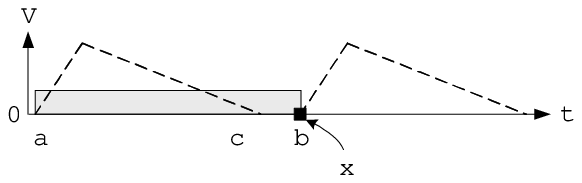}
    \label{f-vertex-inner-null}
}
\caption{Representing envelopes and sigma function using joint vertices.}
\label{f-vertex}
\end{figure*}
%%%%%%%%%%%%%%%%%%%%%%%%%%%%%%%%%%%%%%%%%%%%%%%%%%%%%%%%%%%%%%%%%%%%%%%%%%
%%%%%%%%%%%%%%%%%%%%%%%%%%%%%%%%%%%%%%%%%%%%%%%%%%%%%%%%%%%%%%%%%%%%%%%%%%

%% file: f-example.tex
%%%%%%%%%%%%%%%%%%%%%%%%%%%%%%%%%%%%%%%%%%%%%%%%%%%%%%%%%%%%%%%%%%%%%%%%%%
%%%%%%%%%%%%%%%%%%%%%%%%%%%%%%%%%%%%%%%%%%%%%%%%%%%%%%%%%%%%%%%%%%%%%%%%%%
%% EXAMPLE
%%%%%%%%%%%%%%%%%%%%%%%%%%%%%%%%%%%%%%%%%%%%%%%%%%%%%%%%%%%%%%%%%%%%%%%%%%
%%%%%%%%%%%%%%%%%%%%%%%%%%%%%%%%%%%%%%%%%%%%%%%%%%%%%%%%%%%%%%%%%%%%%%%%%%
\begin{figure*}
\centering
\psfrag{t}{$t$}
\psfrag{0}{\scriptsize $0$}
\psfrag{1}{\scriptsize $1$}
\psfrag{2}{\scriptsize $2$}
\psfrag{3}{\scriptsize $3$}
\psfrag{4}{\scriptsize $4$}
\psfrag{5}{\scriptsize $5$}
\psfrag{6}{\scriptsize $6$}
\psfrag{7}{\scriptsize $7$}
\psfrag{8}{\scriptsize $8$}
\psfrag{9}{\scriptsize $9$}
\psfrag{V}{$V$}
\psfrag{z}{\scriptsize $-1$}
\psfrag{y}{\scriptsize $-2$}
\psfrag{x}{\scriptsize $-3$}
\psfrag{w}{\scriptsize $-4$}
\psfrag{a}{$v_{11}$}
\psfrag{b}{$v_{12}$}
\psfrag{c}{$v_{13}$}
\psfrag{d}{$v_{14}$}
\psfrag{e}{$v_{21}$}
\psfrag{f}{$v_{22}$}
\psfrag{g}{$v_{23}$}
\psfrag{h}{$v_{1}$}
\psfrag{i}{$v_{2}$}
\psfrag{j}{$v_{3}$}
\psfrag{k}{$v_{4}$}
\psfrag{l}{$v_{5}$}
\psfrag{m}{$v_{6}$}
\psfrag{n}{$s_L$}
\psfrag{o}{$s_R$}
\psfrag{H}{\small $f_1(t-s_1)$: Attacker 1}
\psfrag{I}{\small $f_2(t-s_2)$: Attacker 2}
\psfrag{J}{\small $t^*$}
\psfrag{p}{\!\!\!Attacker 1}
\psfrag{q}{\!\!\!Attacker 2}
\psfrag{P}{\small \!$E_{\text{outer}}(t;\mathbf{\Theta}_1)$}
\psfrag{Q}{\small \!$E_{\text{inner}}(t;\mathbf{\Theta}_2)$}
\psfrag{M}{\small ~$\mathbf{\Theta}_1 = (2,3,2,1,3)$}
\psfrag{N}{\small ~$\mathbf{\Theta}_2 = (3,6,-3,0,2)$}
\psfrag{r}{\!\!Sum}
\psfrag{R}{\small $\text{Sum} = E_{\text{max}}(t)$}
\psfrag{S}{\small $E_{\text{min}}(t)$}
\psfrag{T}{\small $f_1(t-s_1^*) + f_2(t-s_2^*)$}
\psfrag{W}{\small $W^*$}
\psfrag{s}{(a)}
\psfrag{u}{(b)}
\psfrag{v}{(c)}
\psfrag{F}{(d)}
\psfrag{G}{(e)}
\psfrag{Z}{\scriptsize $\mathbf{0}^*$}
\psfrag{A}{\scriptsize $\mathbf{1}^*$}
\psfrag{B}{\scriptsize $\mathbf{2}^*$}
\psfrag{C}{\scriptsize $\mathbf{3}^*$}
\psfrag{Y}{\scriptsize $\mathbf{-1}^*$}
\psfrag{D}{Master vertex list}
\psfrag{E}{Attribute tuple chain}
\includegraphics[width=\linewidth]{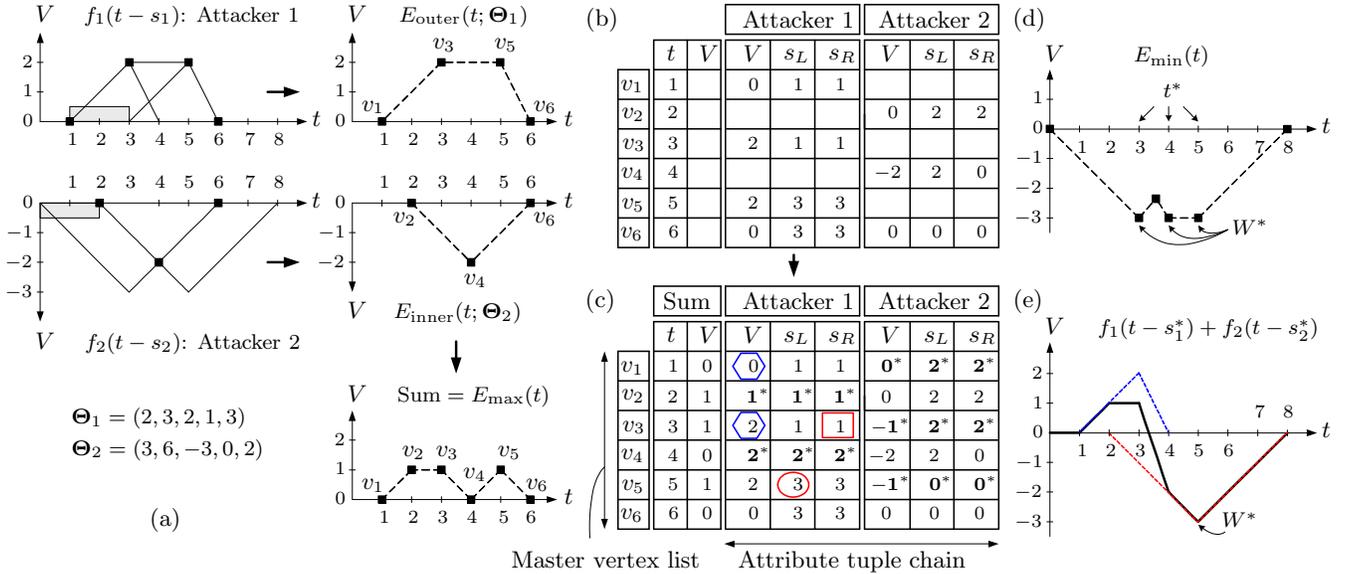}
\caption{Example ($W^*=-3, s_1^*=1, s_2^*=2$).}
\label{f-example}
%in (d), (3.667, -2.33)
\end{figure*}
%%%%%%%%%%%%%%%%%%%%%%%%%%%%%%%%%%%%%%%%%%%%%%%%%%%%%%%%%%%%%%%%%%%%%%%%%%
%%%%%%%%%%%%%%%%%%%%%%%%%%%%%%%%%%%%%%%%%%%%%%%%%%%%%%%%%%%%%%%%%%%%%%%%%%

%% file: exp.tex
FRAME was implemented with C++ and was tested with various synthetic and real circuits. Experiments were performed on a 64-bit 2.66-GHz Linux machine with 8-GB memory. For comparison, we used a commercial static noise analyzer that employs conventional pruning techniques~\cite{chai03,chen01,xiao00,xiao00b,chen99,glebov02acm, arun01,xiao01}. Throughout the experiments, FRAME remained conservative and did not produce optimistic noise figures. The runtime overhead was less than 1\% in all test cases.

%

%The attackers of this circuit form a chain and couple with the victim net by three capacitors.

\begin{figure}[!t]
\centering
\scriptsize
%\subfigure[SPICE waveform at victim (peak: 0.336V)] {
%    \psfrag{t}{$(t)$}
%    \psfrag{0}{$0$}
%    \psfrag{V}{$\!\!\!\!\!\!V$}
%    \psfrag{3}{\!\!\!3ns}
%    \psfrag{4}{\!\!\!4ns}
%    \psfrag{5}{\!\!\!5ns}
%    \psfrag{6}{\!\!\!6ns}
%    \psfrag{a}{\!\!\!\!\!\!0.1}
%    \psfrag{b}{\!\!\!\!\!\!0.2}
%    \psfrag{c}{\!\!\!\!\!\!0.3}
%    \psfrag{d}{\!\!\!\!\!\!0.4}
%    \psfrag{x}{\!\!\!\!\!\!-0.1}
%    \psfrag{e}{}
%    \psfrag{f}{}
%    \psfrag{g}{}
%    \includegraphics[width=0.91\linewidth]{../figure/nsoh-1}
%    \label{f-exp-nsoh-1}
%}
%\subfigure[Comparison] {
%    \psfrag{t}{$(t)$}
%    \psfrag{0}{$0$}
%    \psfrag{V}{$\!\!\!\!\!V$}
%    \psfrag{4}{\!\!\!4ns}
%    \psfrag{8}{\!\!\!8ns}
%    \psfrag{6}{\!\!\!6ns}
%    \psfrag{a}{\!\!\!0.1}
%    \psfrag{b}{\!\!\!0.2}
%    \psfrag{c}{\!\!\!0.3}
%    \psfrag{d}{\!\!\!0.4}
%    \psfrag{m}{\textbf{Conventional pruning}}
%    \psfrag{p}{\!\!\!\!\!\!\!\!\textbf{Proposed}}
%    \includegraphics[width=0.9\linewidth]{../figure/nsoh-2}
%    \label{f-exp-nsoh-1}
%}
\subfigure[Example] {
    \psfrag{e}{100 f$F$}
    \psfrag{f}{50 f$F$}
    \psfrag{g}{200 f$F$}
    \includegraphics[width=0.9\linewidth]{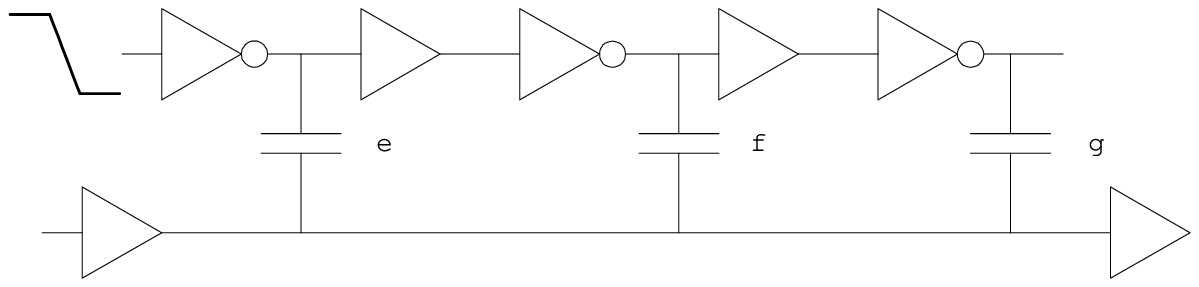}
    \label{f-exp-nsoh-1}
}
\subfigure[Accuracy comparison]{
    \psfrag{t}{\!\!\!\!\!\!\!\!$t$ (ns)}
    \psfrag{v}{$V$}
    \psfrag{a}{\tiny \!\!\!\!\!\!0.1}
    \psfrag{b}{\tiny \!\!\!\!\!\!0.2}
    \psfrag{c}{\tiny \!\!\!\!\!\!0.3}
    \psfrag{d}{\tiny \!\!\!\!\!\!0.4}
    \psfrag{g}[4]{\tiny 3}
    \psfrag{h}[4]{\tiny 5}
    \psfrag{i}[4]{\tiny 7}
    \psfrag{k}[4]{\tiny 9}
    \psfrag{p}{Conventional pruning}
    \psfrag{s}{SPICE}
    \psfrag{f}{FRAME}
    \includegraphics[height=2.25in, width=0.95\linewidth]{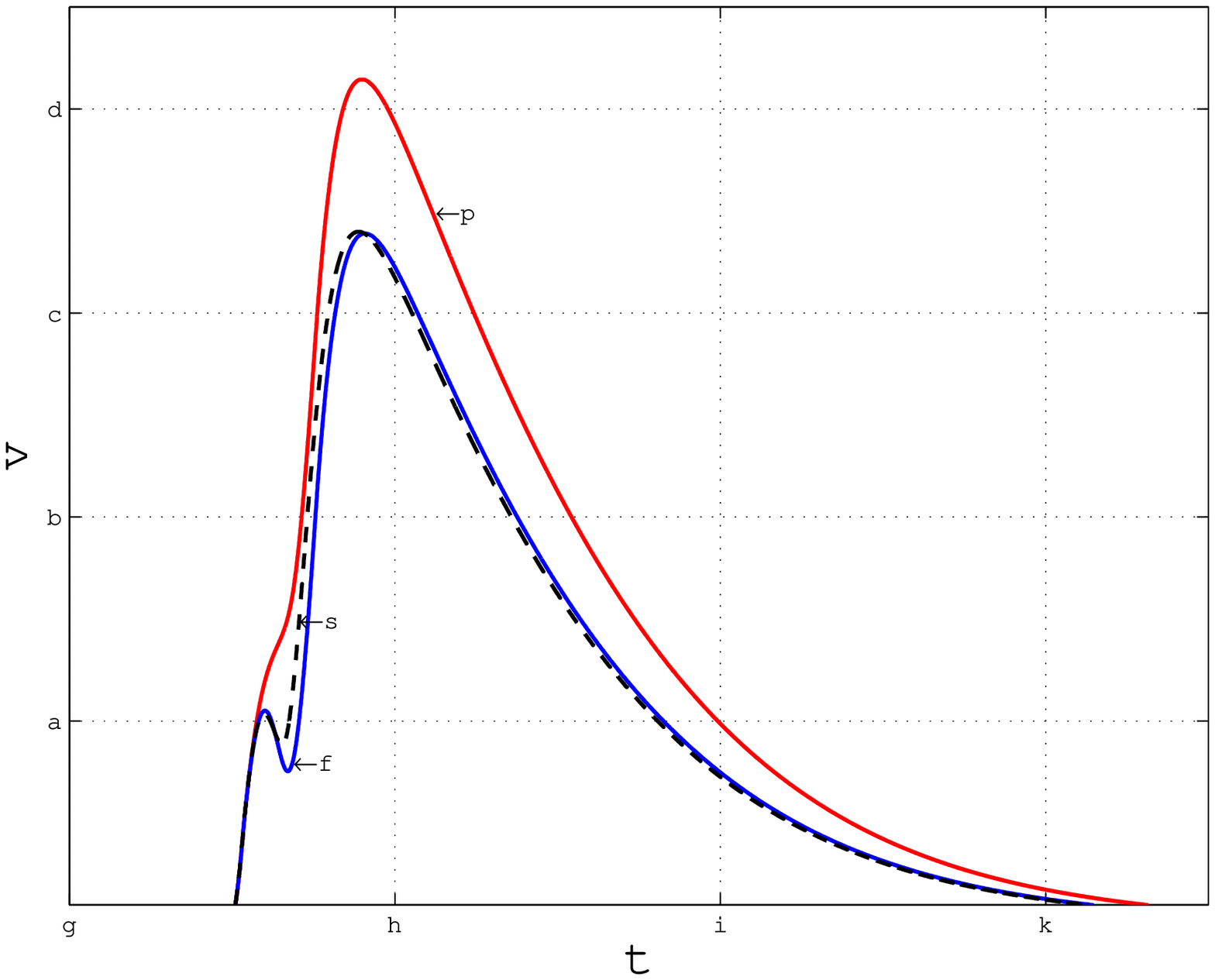}
    \label{f-spice-ex}
}
%\subfigure[] {
%    \psfrag{t}{$(t)$}
%    \psfrag{0}{$0$}
%    \psfrag{V}{$\!\!\!(V)$}
%    \psfrag{4}{4ns}
%    \psfrag{8}{8ns}
%    \psfrag{6}{6ns}
%    \psfrag{a}{\!\!\!0.1}
%    \psfrag{b}{\!\!\!0.2}
%    \psfrag{c}{\!\!\!0.3}
%    \psfrag{d}{\!\!\!0.4}
%    \includegraphics[width=0.85\linewidth]{../figure/nsoh-3}
%    \label{f-exp-nsoh-1}
%}
\caption{Pessimism reduction by FRAME.}
\label{f-exp-pess-reduction}
\end{figure}

We first use a synthetic circuit shown in Fig.~\ref{f-exp-pess-reduction}(a) to explain the pessimism reduction mechanism FRAME relies on. To generate reference data, SPICE is invoked (without breaking the chain) for simulating the victim waveform when the chain head falls. Using the timing windows and triangular parameters obtained through STA, FRAME predicts the alignment of individual attackers, while the commercial tool we used performs pruning. The result is shown in Fig.~\ref{f-exp-pess-reduction}(b), which compares the accuracy of FRAME and the pruning-based commercial tool with respect to SPICE. The peak voltage SPICE, FRAME and the competing tool report are 0.3390V, 0.3398V and 0.4144V, respectively. FRAME thus reduces pessimism by 18.0\% with respect to the competing tool. The FRAME waveform resembles the SPICE curve more closely, producing 93.6\% smaller mean squared error. This difference originates mainly from the fact that the industrial tool eliminates the second attacker because the direction of the second attacker bump is the opposite to that of the others. In other words, retaining the second attacker cannot be the worst-case since removing it induces larger noise, and the commercial tool prunes it. In contrast, the effect of all attackers are considered by FRAME.

%Compared with the SPICE waveform, the FRAME curve exhibits 93.6\% smaller mean squared error

%SPICE was invoked to apply a falling input to the chain head (without breaking the chain) and to simulate the waveform at the victim net.

%industrial design that implements a network processor working at 1.2V and having approximately 1 millon gates integrated by a 90nm process. %The results on overhead and maintaining conservativeness are not presented:

% Although CONV considers temporal and logical correlations, it prunes out an attacker if doing so produces worse noise than not.

%This improvement is mainly due to the fact that FRAME does not prune attackers simply because retaining them cannot induce the worst case. Instead, FRAME evaluates all attackers rapidly and produces a more realistic waveform of combined attackers.

\begin{figure}[b]
\centering
\scriptsize
\subfigure[Comparing worst-case peaks] {
    \psfrag{c}{\!\!\!\!\!\!\!\!\!\!\!\!\!\!\!\!\!\!\!\!\!Conventional $(V)$}
    \psfrag{f}{\!\!\!\!\!\!\!\!\!\!\!\!\!\!\!FRAME $(V)$}
    \includegraphics[width=0.97\linewidth]{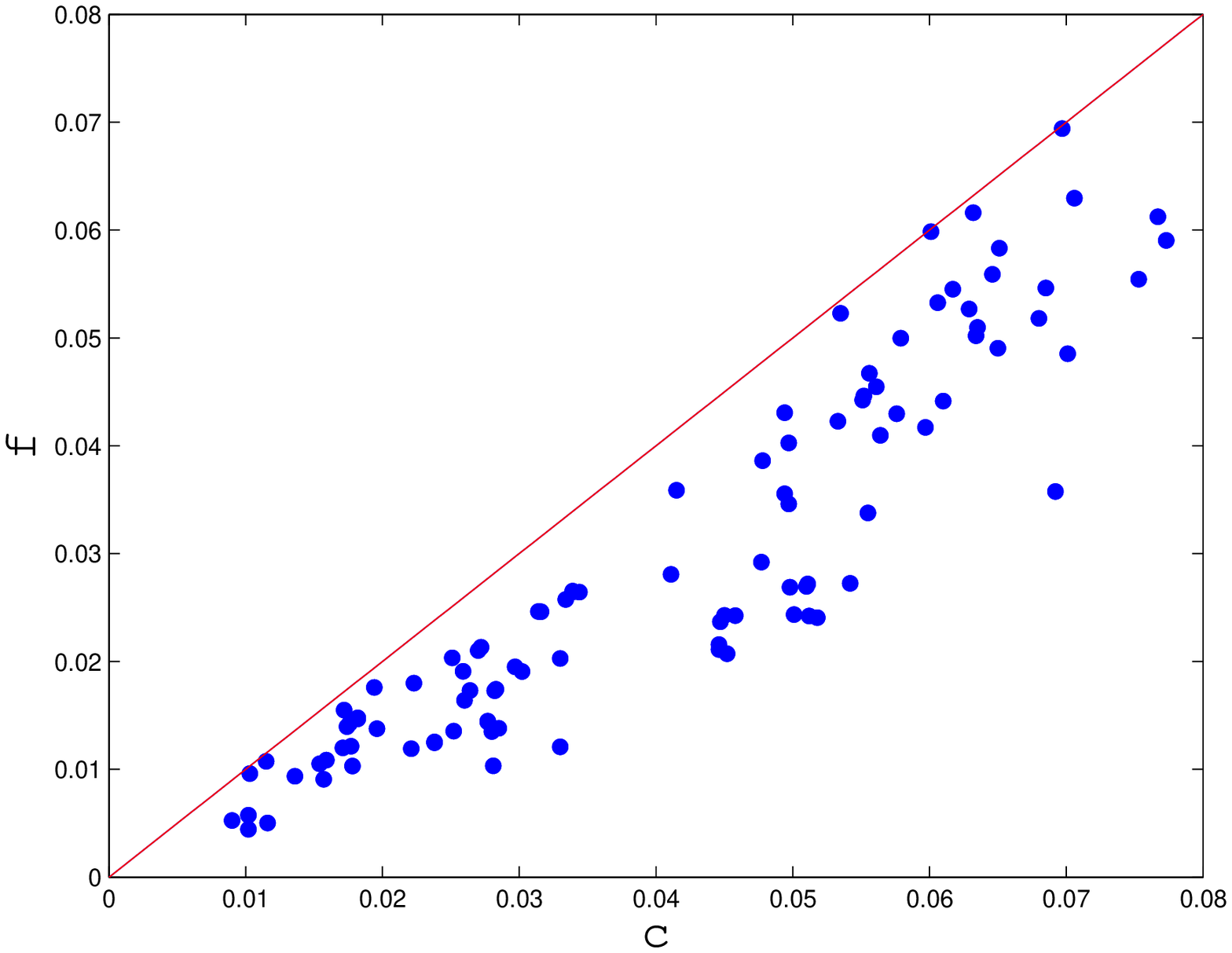}
    \label{f-exp-compare}
}
%\subfigure[Box plot for pessimism reduction (\%)] {
\subfigure[Distribution of reduced pessimism (\%)] {
    \psfrag{p}{$\%$}
    \psfrag{0}{$0$}
    \psfrag{2}{$20$}
    \psfrag{4}{$40$}
    \psfrag{6}{$60$}

%    \psfrag{\%}{$\%$}
    \includegraphics[width=0.8\linewidth]{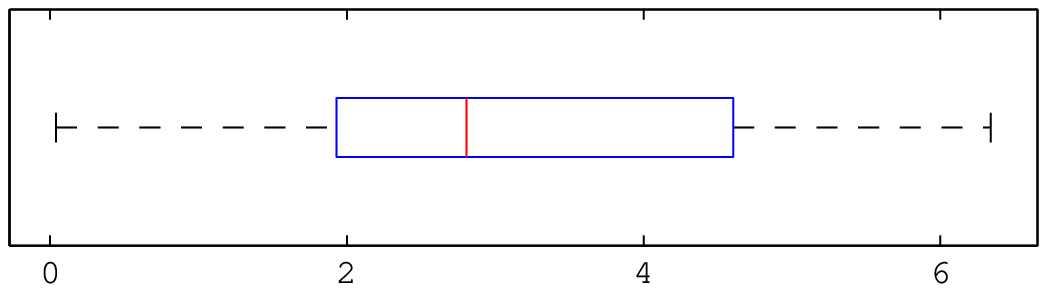}
    \label{f-exp-box}
}
\caption{Comparing worst-case peaks estimated over 201 circuits by FRAME and conventional tool.}
\end{figure}

We extensively tested FRAME with 201 circuits extracted from a 90-nm network processor design with approximately 1 million gates.  Fig.~\ref{f-exp-compare} compares the performance of FRAME with that of the commercial tool in terms of the worst-case peak voltage. For each dot at $(x,y)$, $x$ and $y$ represents the worst-case peak voltages estimated by the commercial tool and FRAME, respectively. FRAME could reduce pessimism significantly, by 30.4\% on average, compared with the pruning-based industrial tool. The distribution of the pessimism reduction amount is shown in Fig.~\ref{f-exp-box}, in which a boxplot is used for showing the median (the line inside the box) and interquartile range~\cite{rice95}. This plot shows that FRAME typically reduces pessimism around 20-45\% for the 201 circuits we tested.

To prove further the statistical significance of our result, we performed the Mann-Whitney-Wilcoxon test~\cite{rice95}, a nonparametric test for assessing whether two sample sets come from the same distribution. According to this test, we could reject the hypothesis that two groups of worst-case peak voltage observations (one from FRAME; the other from the commercial tool) come from the same distribution at the 0.01\% significance level with $p$-value of $2.85\times10^{-10}$. This result indicates that the observed difference in noise figures is statistically significant and meaningful.

%General info: 90 nm, 1.2 V.
%Circuit: slide 1 from the slides I sent to you last time.
%
%1) tlc_vs_real_lc_with_aggr: This is white background version of slide 5 . In this figure,
%  top: TLC version with aggressor chains are broken into individual aggressors. thus, aggressor wave forms are piece wise linear.
%  middle: original logical correlation version where aggressors are connected as one chain. Thus, the second and third aggressor wave forms are smooth real wave forms.
%  bottom: Comparison of TLC (top victim) and original (middle victim). Height are matching well.
%
%2) tlc_and_prune: Compare TLC (result of 1) above) and pruned version where the second aggressor is pruned. Since the second aggressor is omitted, there's no opposite direction aggressor, so the final bump is bigger.
%TLC: 0.33697 mV
%pruned: 0.41436 mV
%
%3) 3_inv_chain: white back ground version of slide 1